\newif\ifColor\Colortrue \Colorfalse
\Colortrue
\PassOptionsToPackage{super,comma,numbers,sort&compress}{natbib}  
\documentclass[preprint, 5p, 8pt, lefttitle]{elsarticle}

\usepackage{xpatch}
\xpatchcmd{\MaketitleBox}{\hrule\vskip12pt}{\vspace{-2\baselineskip}}{}{}% remove first horizontal rule (above abstract)
\xpatchcmd{\MaketitleBox}{\hrule}{}{}{}% remoce second horizonral rule (below keywords)

\usepackage[switch, left]{lineno}
\modulolinenumbers[1]
\usepackage[utf8]{inputenc}
\usepackage{amssymb}
\usepackage{amsmath}
\usepackage{amsfonts}
\usepackage{upgreek}
\usepackage{gensymb}
\usepackage{setspace}
\usepackage{booktabs}
\usepackage[english]{babel}
\usepackage{mathtools}
\usepackage{mhchem}
\usepackage{hyperref}
\usepackage[final]{microtype}
\usepackage{csquotes,lmodern}
\usepackage{subcaption}
\usepackage{etoolbox}
\usepackage{color}
\usepackage{multibib}
\newcites{methods}{References}
\usepackage[labelfont=bf]{caption}
\DeclareCaptionLabelSeparator{thinspace}{\,}
\DeclareCaptionLabelFormat{mylabel}{#1\,#2}

\usepackage{sidecap}
\usepackage{adjustbox}

\makeatletter
\renewenvironment{abstract}{\global\setbox\absbox=\vbox\bgroup
  \hsize=\textwidth\def\baselinestretch{0}%
 \par\unskip\noindent\unskip\ignorespaces}
 {\egroup}

\def\keyword{%
  \def\sep{\unskip, }%
 \def\MSC{\@ifnextchar[{\@MSC}{\@MSC[2000]}}
  \def\@MSC[##1]{\par\leavevmode\hbox {\it ##1~MSC:\space}}%
 \def\PACS{\par\leavevmode\hbox {\it PACS:\space}}%
  \def\JEL{\par\leavevmode\hbox {\it JEL:\space}}%
 \global\setbox\keybox=\vbox\bgroup\hsize=\textwidth
  \normalsize\normalfont\def\baselinestretch{0}
 \parskip\z@
  \noindent\textit{Some important words: }   <--- Edit as necessary
  \raggedright                         % Keywords are not justified.
  \ignorespaces}

\def\ps@pprintTitle{%
     \let\@oddhead\@empty
     \let\@evenhead\@empty
     \def\@oddfoot{\footnotesize\itshape
      \ifx\@journal\@empty  % <--- Edit as necessary
       \else\@journal\fi\hfill\today}%
     \let\@evenfoot\@oddfoot}
     
\makeatother

\begin{document}
\begin{frontmatter}
%High-performance \ce{Fe2VAl} composite thermoelectrics with topological-insulator grain boundary networks
\title{\onehalfspacing\textbf{\fontsize{18}{18}\selectfont{\textsf{Decoupled charge and heat transport for high-performance \ce{Fe2VAl} composite thermoelectrics}}}}

\author[1]{\textsf{\small\textbf{Fabian Garmroudi}}\corref{cor1}}
\author[2]{\textsf{\small\textbf{Illia Serhiienko}}}
\author[1]{\textsf{\small\textbf{Michael Parzer}}}
\author[3]{\textsf{\small\textbf{Sanyukta Ghosh}}}
\author[3]{\textsf{\small\textbf{Pawel Ziolkowski}}}
\author[3]{\textsf{\small\textbf{Gregor Oppitz}}}
\author[4]{\textsf{\small\textbf{Hieu Duy Nguyen}}}
\author[2,5]{\textsf{\small\textbf{Cédric Bourgès}}}
\author[2]{\textsf{\small\textbf{Yuya Hattori}}}
\author[1]{\textsf{\small\textbf{Alexander Riss}}}
\author[1]{\textsf{\small\textbf{Sebastian Steyrer}}}
\author[6]{\textsf{\small\textbf{Gerda Rogl}}}
\author[6]{\textsf{\small\textbf{Peter Rogl}}}
\author[7]{\textsf{\small\textbf{Erhard Schafler}}}
%\author[7]{\textsf{\small\textbf{Taras Parashchuk}}}
%\author[7]{\textsf{\small\textbf{Krzysztof Wojciechowski}}}
\author[4]{\textsf{\small\textbf{Naoyuki Kawamoto}}}
\author[3,8]{\textsf{\small\textbf{Eckhard M\"uller}}}
\author[1]{\textsf{\small\textbf{Ernst Bauer}}}
\author[3,9]{\textsf{\small\textbf{Johannes de Boor}}\corref{cor1}}
\author[2,10]{\textsf{\small\textbf{Takao Mori}}\corref{cor1}}
\cortext[cor1]{Correspondence to:\,f.garmroudi@gmx.at,\\Johannes.deBoor@dlr.de,\,MORI.Takao@nims.go.jp}
\address[1]{\textsf{Institute of Solid State Physics, TU Wien, 1040 Vienna, Austria}}
\address[2]{\textsf{International Center for Materials Nanoarchitectonics (WPI-MANA), National Institute for Materifals Science (NIMS), Tsukuba, Japan}}
\address[3]{\textsf{Institute of Materials Research, German Aeropspace Center (DLR), D-51147 Cologne, Germany}}
\address[4]{\textsf{Center for Basic Research on Materials (CBRM), National Institute for Materifals Science (NIMS), Tsukuba, Japan}}
\address[5]{\textsf{International Center for Young Scientists, National Institute for Materials Science (NIMS), Tsukuba, Japan}}
\address[6]{\textsf{Institute of Materials Chemistry, University of Vienna, 1090 Vienna, Austria}}
\address[7]{\textsf{Faculty of Physics, University of Vienna, 1090 Vienna, Austria}}
\address[8]{\textsf{Institute of Inorganic and Analytical Chemistry, Justus Liebig University Giessen, D-35392 Giessen, Germany}}
\address[9]{\textsf{University of Duisburg-Essen, Faculty of Engineering, Institute of Technology for Nanostructures (NST) and CENIDE, D-47057 Duisburg, Germany}}
\address[10]{\textsf{Graduate School of Pure and Applied Sciences, University of Tsukuba, Tsukuba, Japan}}

\selectlanguage{english}
\journal{submitted to Nature Materials}

\begin{abstract}
\noindent \textsf{\fontsize{10}{10}\selectfont{Decoupling charge and heat transport is essential for optimizing thermoelectric materials. Strategies to inhibit lattice-driven heat transport, however, also compromise carrier mobility, limiting the performance of most thermoelectrics, including \ce{Fe2VAl} Heusler compounds. Here, we demonstrate an innovative approach, which bypasses this tradeoff: via liquid-phase sintering, we incorporate the archetypal topological insulator \ce{Bi_{1-x}Sb_{x}} between \ce{Fe2V_{0.95}Ta_{0.1}Al_{0.95}} grains. Structural investigations alongside extensive thermoelectric and magneto-transport measurements reveal distinct modifications in the microstructure, and a reduced lattice thermal conductivity and enhanced carrier mobility are simultaneously found. This yields a huge performance boost -- far beyond the effective-medium limit -- and results in one of the highest figure of merits among both half- and full-Heusler compounds, $z\approx 1.6\times 10^{-3}\,$K$^{-1}$ ($zT\approx 0.5$) at 295\,K. Our findings highlight the potential of secondary phases to decouple charge and heat transport and call for more advanced theoretical studies of multiphase composites.}}
\end{abstract}

\end{frontmatter}

%\begin{linenumbers}

\noindent Given the increasing global demand for efficient energy utilization, thermoelectrics (TEs) present a promising solution as they can harvest decentralized waste heat sources or function as Peltier coolers, e.g. for thermal management applications \cite{rowe2018thermoelectrics,
shi2020advanced}. The conversion efficiency of TE devices depends on the hot- and cold-side temperatures and a material-dependent figure of merit, $z \propto \mu_\text{W}/\kappa_\text{L}$. The highest achievable $z$ in a semiconductor with optimized carrier concentration is determined by the weighted carrier mobility $\mu_\text{W}$ of electrons or holes, which should be maximized, and by the lattice thermal conductivity $\kappa_\text{L}$, which should be minimized \cite{zevalkink2018practical,snyder2020weighted}. The inherent tradeoff between $\mu_\text{W}$ and $\kappa_\text{L}$ presents one of the most formidable challenges in the design and optimization of TE materials, requiring the decoupling of charge and heat transport, that is, the realization of a phonon-glass electron-crystal concept.

%The greatest potential for thermoelectrics lies between 300 and 400\,K, where most waste heat is released into the environment. 
Since the mid-20$^\text{th}$ century, \ce{Bi2Te3}-based systems have been the gold standard for TEs operating near room temperature, and currently, they remain as the only commercially available option \cite{goldsmid1958electrical,witting2019thermoelectric}. However, the scarcity of tellurium, along with the brittle nature and poor mechanical properties of these materials limits their widespread use in everyday life and industrial applications. Therefore, it is crucial to explore alternatives that offer competitive performance and overcome the challenges related to \ce{Bi2Te3}.

For $n$-type materials, cost-effective \ce{Mg3(Bi,Sb)2} Zintl compounds have been considered the hottest candidates as they exhibit very high $z$ \cite{mao2019high,liu2021demonstration,
liu2022maximizing}. However, these materials, especially the Bi-rich alloys with attractive near-room temperature properties, suffer from poor chemical stability and degrade rapidly when exposed to air, presenting a significant and unresolved dilemma for practical applications. 

On the other hand, Heusler compounds based on \ce{Fe2VAl}, the focus of this study, exhibit excellent chemical and mechanical stability. These materials are also composed of earth-abundant, inexpensive elements with great recyclability  \cite{roadmap2024} -- sustainability aspects that are becoming increasingly important worldwide, and particularly within the EU. Moreover, \ce{Fe2VAl} full-Heuslers display oustanding electronic transport properties, with weighted mobilities that are comparable to or even greater than those of other state-of-the-art TEs \cite{anand2020thermoelectric}. Yet, their intrinsically large $\kappa_\text{L}$ limits their potential as TE materials \cite{alleno2018review}. Consequently, previous studies have primarily focused on reducing $\kappa_\text{L}$ by substituting heavy elements to scatter heat-carrying phonons \cite{mikami2012thermoelectric,
renard2014thermoelectric,hinterleitner2020stoichiometric}, reducing the dimensionality through thin-film deposition \cite{furuta2014fe,hiroi2016thickness,
bourgault2023improved}, or reducing the grain size \cite{mikami2008synthesis,masuda2018effect,alleno2018review}. Although these strategies have resulted in enhancements of $z$, the overall performance remains a significant bottleneck and is too low for most practical applications.

\begin{figure*}[tbh]
\newcommand{\setwidth}{0.45}
			\centering
			\hspace*{0cm}
		\includegraphics[width=0.95\textwidth]{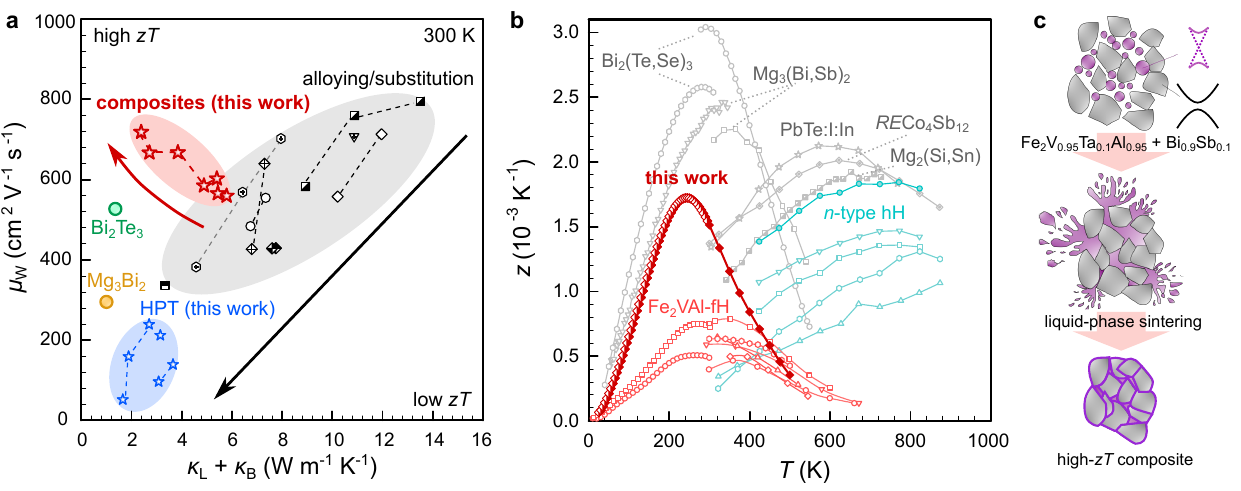}
	\caption{\textbf{$\vert$\,Boosting thermoelectric performance in Heusler compounds by decoupled charge and heat transport.} \textbf{a}, Tradeoff between weighted mobility and lattice thermal conductivity (plus bipolar term $\kappa_\text{B}$) in \ce{Fe2VAl}-based Heusler compounds at room temperature \cite{nishino2006thermal,mikami2012thermoelectric,miyazaki2013thermoelectric,
	takeuchi2013effect,renard2014thermoelectric}. Data for state-of-the-art $n$-type \ce{Bi2Te3}- and \ce{Mg3Bi2}-based systems \cite{witting2019thermoelectric,mao2019high} at 300\,K are shown for comparison. Composites in this work rae found to bypass this tradeoff. \textbf{b}, Temperature-dependent figure of merit of the best composite (\ce{Fe2V_{0.95}Ta_{0.1}Al_{0.95}}+50\,vol.\% \ce{Bi_{0.9}Sb_{0.1}} added), compared to other high-performance $n$-type thermoelectrics \cite{rogl2017v,chauhan2018vanadium,
	chauhan2019compositional,sankhla2018mechanical,
	meng2017grain,zhang2018deep}. \textbf{c}, Schematic synthesis of composites via liquid-phase sintering.} 
	\label{Fig1}
\end{figure*}

In this study, we demonstrate that by incorporating chemically and structurally distinct \ce{Bi_{1-x}Sb_{x}} at the grain boundaries, charge and heat transport can be decoupled, resulting in a reduction of $\kappa_\text{L}$, and simultaneously, in an unexpected increase of $\mu_\text{W}$ (Fig.\,\ref{Fig1}a). Consequently, the figure of merit is boosted by more than a factor of two, up to $z \approx 1.6\times 10^{-3}\,$K$^{-1}$ ($zT\approx 0.5$) at room temperature, representing one of the largest values hitherto reported for $n$-type half- and full-Heusler compounds (see Fig.\,\ref{Fig1}b).

\vspace*{-0.2cm}
\section*{Decoupling charge and heat transport in \ce{Fe2VAl}}
\vspace*{-0.1cm}
\noindent The lattice thermal conductivity of \ce{Fe2VAl} Heusler compounds is intrinsically large, $\kappa_\text{L}\approx 27\,$W\,m$^{-1}$\,K$^{-1}$ at 300\,K \cite{nishino2006thermal}, which can be mainly attributed to a lack of structural and chemical bonding complexity as well as the absence of heavy elements, leading to high sound velocities. Upon alloying, $\kappa_\text{L}$ can be drastically reduced down to 10\,W\,m$^{-1}$\,K$^{-1}$ in \ce{Fe2VAl_{1-x}Si_{x}} \cite{nishino2006thermal}, 7\,W\,m$^{-1}$\,K$^{-1}$ in \ce{Fe2VAl_{1-x}Ge_{x}} \cite{nishino2006thermal}, and by substituting heavy 5$d$ elements, further down to 5\,W\,m$^{-1}$\,K$^{-1}$ in \ce{Fe2VTa_{x}Al_{1-x}} \cite{renard2014thermoelectric} and 4\,W\,m$^{-1}$\,K$^{-1}$ in \ce{Fe2V_{1-x}W_{x}Al} \cite{hinterleitner2020stoichiometric}. As a downside, the very same point defects, which effectively inhibit heat transport by high-frequency phonons, also strongly scatter charge carriers. This is particularly true for the 5$d$ elements like Ta and W, which are substituted for V atoms. Since V-$d$ states dominate the electronic states of the conduction band, introducing substitutional disorder at the V site results in intense electronic scattering. We gathered TE property data from various substitution studies \cite{nishino2006thermal,mikami2012thermoelectric,miyazaki2013thermoelectric,
	takeuchi2013effect,renard2014thermoelectric} and calculated $\mu_\text{W}$ \cite{snyder2020weighted}. A strong tradeoff relationship between $\mu_\text{W}$ and $\kappa_\text{L}$ is obvious from Fig.\,\ref{Fig1}a (black symbols).

\begin{figure*}[t!]
\newcommand{\setwidth}{0.45}
			\centering
			\hspace*{0cm}
		\includegraphics[width=0.95\textwidth]{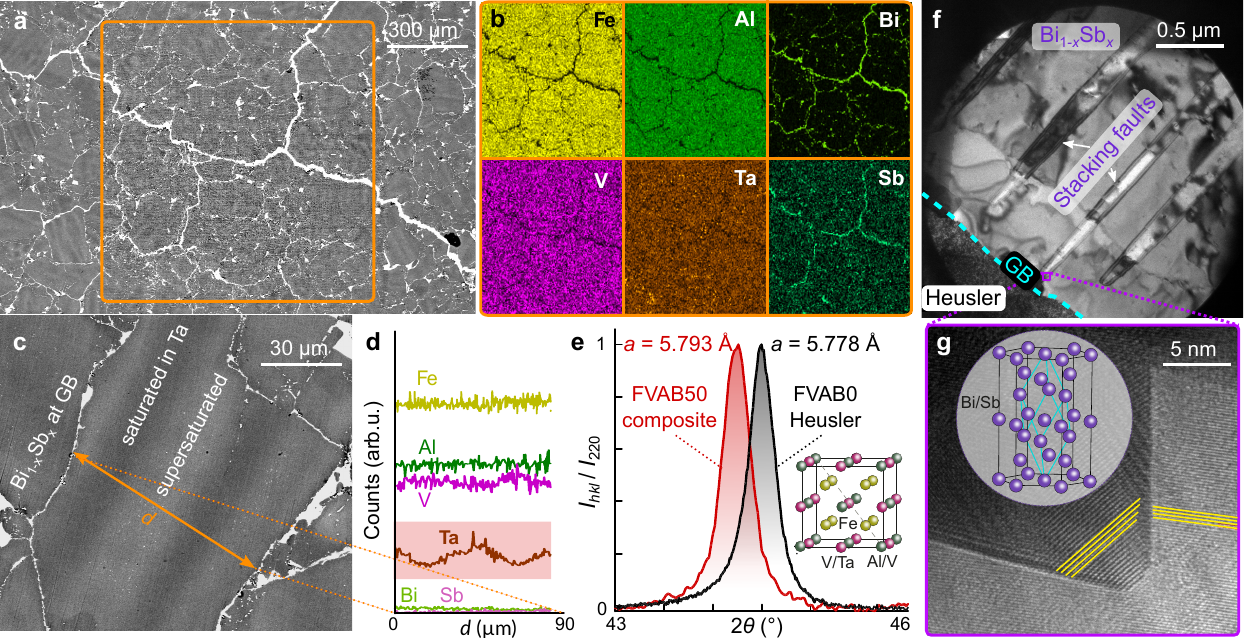}
	\caption{\textbf{$\vert$\,Microstructure evolution in \ce{Fe2V_{0.95}Ta_{0.1}Al_{0.95}} Heusler compounds upon incorporating \ce{Bi_{1-x}Sb_{x}}.} \textbf{a}, Microstructure of FVAB50 composite, where the majority of GBs are filled with Bi-Sb. \textbf{b}, EDX analyses reveal that Bi and Sb are found at the GBs, while Fe, V, Al and Ta are almost exclusively distributed within the grains. \textbf{c}, BS-SEM image of a Heusler grain with periodic contrast variations, surrounded by Bi-Sb. \textbf{d}, EDX line scan along the Heusler grain shown in \textbf{c}. \textbf{e}, Comparison of normalized X-ray diffraction peaks of the (220) plane of \ce{Fe2V_{0.95}Ta_{0.1}Al_{0.95}} and FVAB50 composite. \textbf{f}, Bright-field TEM image of the GB with ladder-like nanostructure arrays of stacking faults and \textbf{g}, high-magnification image near stacking-faults. Insets in \textbf{e} and \textbf{g} show Heusler and Bi-Sb unit cells, respectively.}  
	\label{Fig2}
\end{figure*}

Aside from introducing point defects, $\kappa_\text{L}$ can be suppressed by reducing the grain size $d$ and several studies attempted to enhance $z$ by grain size reduction, e.g., via ball milling \cite{mikami2008synthesis,mikami2012thermoelectric} or high-pressure torsion (HPT) \cite{masuda2018effect,fukuta2022improving}, yielding $d \approx 100\,$nm. Employing HPT, Fukuta et al. recently reported very low values of $\kappa_\text{L}$ down to 1.3\, W\,m$^{-1}$\,K$^{-1}$ in \ce{Fe2V_{0.98}Ta_{0.1}Al_{0.92}} at 350\,K and $zT$ up to 0.37 at 400\,K ($z\approx 0.93\times 10^{-3}$\,K$^{-1}$) \cite{fukuta2022improving}. These remarkable findings motivated us to (i) reproduce them and (ii) apply HPT to a variety of different samples with optimized compositions. The results of these endeavors are summarized in the Supplementary Information (SI). While $\kappa_\text{L}$ could indeed be dramatically reduced down to $<2$\, W\,m$^{-1}$\,K$^{-1}$, we concomitantly observed a huge deterioration of electronic transport in all cases (see Fig.\,S1-S2 and blue symbols in Fig.\,\ref{Fig1}a), resulting in no enhancement of $zT$ (Fig.\,S3). Similar observations have been made, e.g., for \ce{Mg3(Bi,Sb)2} \cite{wood2019improvement,luo2021nb} and various half-Heuslers, where reducing grain size comes at a cost of reducing $\mu_\text{W}$ \cite{qiu2019grain,bueno2023grain}. The discrepancy between our results and previous ones \cite{fukuta2022improving} suggests that setup-specific conditions during HPT are generally very important, complicating reproducibility and upscale production.    

Instead, we have devised a different approach wherein chemically and structurally distinct \ce{Bi_{1-x}Sb_{x}} is incorporated as a secondary phase between the Heusler grains. Fig.\,\ref{Fig1}c outlines the synthesis procedure. The starting materials were first synthesized using an induction melting furnace and then hand-ground into a fine powder. The powders were mixed in various ratios (5\,--\,50\,vol.\% \ce{Bi_{0.9}Sb_{0.1}}), and sintered at 1373\,K. The much lower melting point of \ce{Bi_{0.9}Sb_{0.1}} causes excess liquid to be expelled during sintering. The retention of \ce{Bi_{1-x}Sb_{x}} in the composite depends on the particle size of the Heusler phase and the amount of \ce{Bi_{0.9}Sb_{0.1}} used. Backscattered scanning electron microscopy (BS-SEM) shows that up to $\approx 30$\,vol.\%, \ce{Bi_{0.9}Sb_{0.1}} fills only the triple junctions of Heusler grains, while 50\,vol.\% \ce{Bi_{0.9}Sb_{0.1}} allows the liquid phase to wet and coat all grains as a grain boundary (GB) phase (Fig.\,S10). This produces highly dense (\ce{Fe2V_{0.95}Ta_{0.1}Al_{0.95}}\,+\,\ce{Bi_{0.9}Sb_{0.1}}) composites, referred to as FVAB$X$, with $X$ indicating the \ce{Bi_{0.9}Sb_{0.1}} volume percentage added before sintering.  

The $\mu_\text{W}$ versus $\kappa_\text{L}$ trend (see Fig.\,\ref{Fig1}a) for the composite samples is unusual and cardinally different from other approaches, like alloying or grain size reduction via HPT. Moreover, the exceptionally high $\mu_\text{W}$, in spite of the suppressed $\kappa_\text{L}$, signifies a decoupling of charge and lattice-driven heat transport. In the following, we present investigations of the microstructure of these materials alongside local microscale probing of the Seebeck coefficient $S$. Finally, we show and discuss experimental results from extensive TE and magneto-transport measurements carried out in a broad range of temperatures and magnetic fields.

\vspace*{-0.2cm}
\section*{Structural modifications in FVAB$X$ composites}
\vspace*{-0.1cm}
\noindent The structural properties of the sintered samples were investigated via scanning and transmission electron microscopy (TEM), energy-dispersive X-ray spectroscopy (EDX) and X-ray diffraction (XRD). Fig.\,S4 shows BS-SEM images of \ce{Fe2V_{0.95}Ta_{0.1}Al_{0.95}} sintered at 1373\,K without the addition of \ce{Bi_{0.9}Sb_{0.1}}. Throughout the whole sample, nanoscale precipitates of a secondary Ta-rich phase are clearly noticeable at the GBs. This is in agreement with the previously established low solubility limit of Ta, $x=0.07$ in \ce{Fe2V_{1-x}Ta_xAl}\,\cite{garmroudi2021solubility}. Apart from that, the microstructure displays a very homogeneous phase distribution without any variations in the composition.

Fig.\,\ref{Fig2}a shows a low-magnification image of the microstructure of the FVAB50 composite, with the best TE properties. A uniform distribution of \ce{Bi_{1-x}Sb_{x}} along the GBs is obvious and confirmed by compositional mapping using EDX (Fig.\,\ref{Fig2}b). Additionally, we find that the segregation of \ce{Bi_{1-x}Sb_{x}} along the GBs goes hand in hand with two changes in the microstructure: (i) strong suppression of nanoscale Ta-rich precipitates at the GBs, (ii) diffuse brightness variations within the grains. Both these structural changes suggest an enhanced solubility limit of Ta, when \ce{Bi_{1-x}Sb_{x}} is incorporated as a GB network during the liquid-phase sintering. This is confirmed by EDX line scans (Fig.\,\ref{Fig2}d) across the grain, revealing periodic fluctuations in the Ta and V concentration, and XRD, revealing an increase of the lattice parameter of the Heusler phase (Fig.\,\ref{Fig2}e) as larger Ta atoms are substituted. In Fig.\,\ref{Fig2}f,g, we focus on \ce{Bi_{1-x}Sb_{x}}. Since \ce{Fe2VAl} and \ce{Bi_{1-x}Sb_{x}} are chemically and structurally distinct, there exists a well-defined GB without apparent inter-diffusion. We find that \ce{Bi_{1-x}Sb_{x}}, when embedded between Heusler grains, displays a peculiar ladder-like nanostructure with arrays of stacking fault defects. Moreover, detailed EDX analyses of different samples indicates that the Bi:Sb ratio fluctuates and that the Sb concentration is well above the nominal one in our high-performance FVAB50 composite (Fig.\,S9 and S11). 

\begin{figure*}[t]
\newcommand{\setwidth}{0.45}
			\centering
			\hspace*{0cm}
		\includegraphics[width=0.85\textwidth]{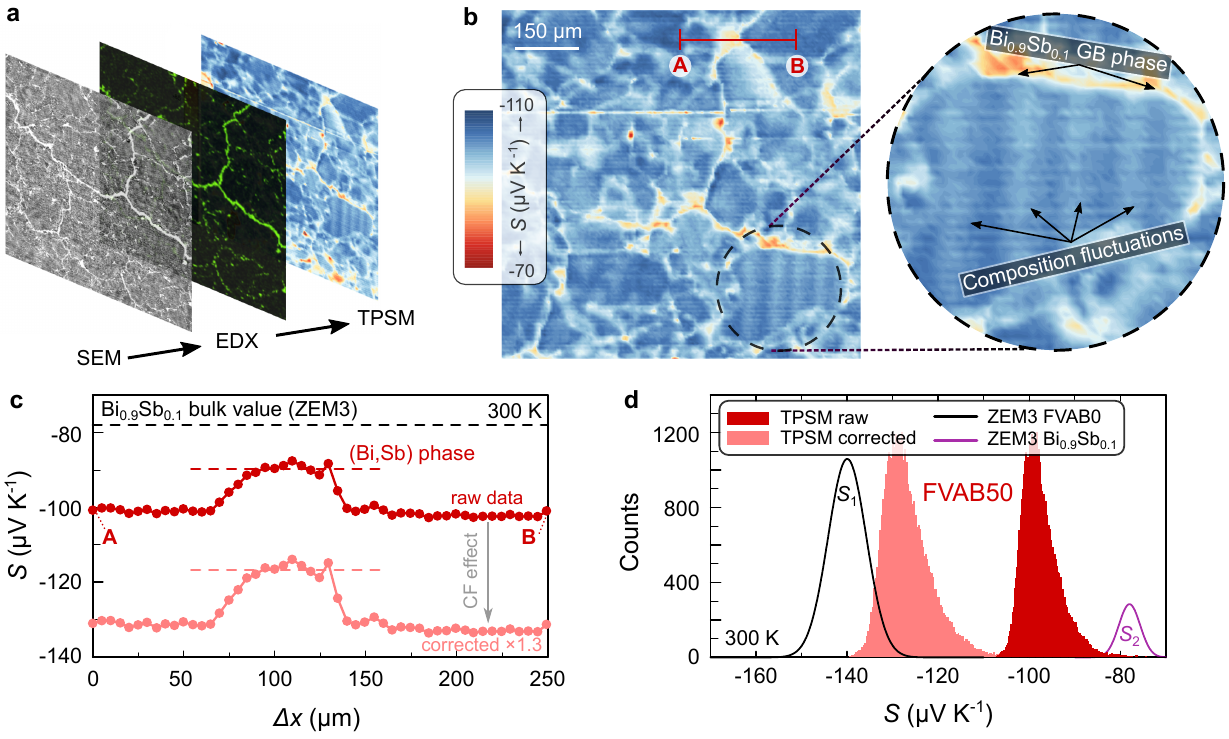}
	\caption{\textbf{$\vert$\,Local investigation of electronic transport in (\ce{Fe2V_{0.95}Ta_{0.1}Al_{0.95}}\,+\,\ce{Bi_{0.9}Sb_{0.1}}) composites.} \textbf{a}, Microstructure, composition and local transport probing in the same area (orange square in Fig.\,2a). \textbf{b}, \textbf{c}, Transient potential Seebeck microprobe (TPSM) mapping of FVAB50 at room temperature. Ta-enriched regions inside the Heusler grains and \ce{Bi_{1-x}Sb_{x}} at the GBs display smaller $S$ than the remaining part of the matrix. \textbf{c}, TPSM line scans across distance marked in \textbf{b}. \textbf{d}, Histogram from TPSM mapping. Solid lines are normal distributions centered around $S_1=-140\,\mu$V$^{-1}$K$^{-1}$ and $S_2=-78\,\mu$V$^{-1}$K$^{-1}$, the bulk values for \ce{Fe2V_{0.95}Ta_{0.1}Al_{0.95}} and \ce{Bi_{0.9}Sb_{0.1}}, respectively.} 
	\label{Fig3}
\end{figure*}

\vspace*{-0.2cm}
\section*{Breaking the effective-medium limit}
\vspace*{-0.1cm}
\noindent The pivotal role of understanding and investigating TE transport across grain boundaries is increasingly recognized \cite{kuo2018grain,wood2019improvement,ghosh2023towards}. To draw a connection between microstructure and electronic transport we employed a transient potential Seebeck microprobe (TPSM), with local property investigations performed on the same rectangular area of the sample (Fig.\,\ref{Fig3}a). Fig.\,\ref{Fig3}b shows a map of the locally determined $S$ with a spatial resolution of 3\,--\,5 microns. The results obtained are in excellent agreement with structural investigations revealing a rich and complex microstructure consisting of Heusler grains and a \ce{Bi_{1-x}Sb_{x}} GB network. Interestingly, TPSM measurements suggest that \ce{Bi_{1-x}Sb_{x}} exhibits a larger $S$ as a secondary phase compared to its bulk form. This is emphasized by looking at line scans across Heusler grains. The plateau in Fig.\,\ref{Fig3}c refers to the value of \ce{Bi_{1-x}Sb_{x}} within the composite, which is significantly higher than its bulk value, especially considering that TPSM measurements typically underestimate $S$ by at least 20\,--\,30\,\% due to the cold finger effect. This enhancement, which exceeds the highest $S$ at 300\,K in the entire composition range of polycrystalline \ce{Bi_{1-x}Sb_x} \cite{doroshenko2016thermoelectric}, is surprising, given the near-complete immiscibility between \ce{Bi_{1-x}Sb_{x}} and the Heusler phase. 

\begin{figure*}[t!]
\newcommand{\setwidth}{0.45}
			\centering
			\hspace*{0cm}
		\includegraphics[width=0.8\textwidth]{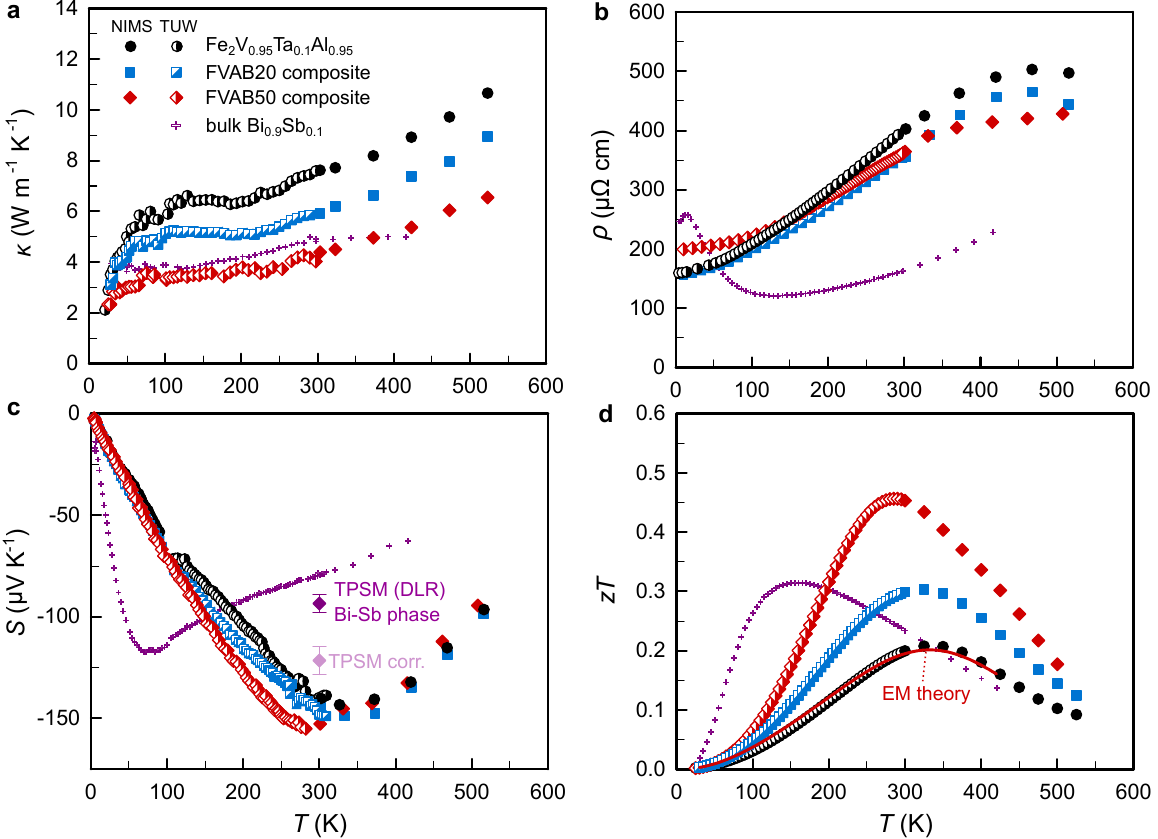}
	\caption{\textbf{$\vert$\,Thermoelectric properties of (\ce{Fe2V_{0.95}Ta_{0.1}Al_{0.95}}\,+\,\ce{Bi_{0.9}Sb_{0.1}}) composites.} \textbf{a}, Temperature-dependent thermal conductivity, \textbf{b}, electrical resistivity, \textbf{c}, Seebeck coefficient, and \textbf{d}, dimensionless figure of merit $zT$ of \ce{Fe2V_{0.95}Ta_{0.1}Al_{0.95}} composites with 20 and 50\,vol.\% \ce{Bi_{0.9}Sb_{0.1}} added before sintering (FVAB20, FVAB50) compared to pristine \ce{Fe2V_{0.95}Ta_{0.1}Al_{0.95}} and \ce{Bi_{0.9}Sb_{0.1}}. The error bars in \textbf{c} indicate the statistical variation of the Bi-Sb phase corresponding to the area shown in Fig.\,\ref{Fig3}b. The red solid line in \textbf{d} represents a calculation based on effective-medium theory for FVAB50, using a volume fraction of $\approx 6$\,vol.\% \ce{Bi_{0.9}Sb_{0.1}}, determined by EDX.} 
	\label{Fig4}
\end{figure*}

Fig.\,\ref{Fig3}d shows the distribution histogram of the measured Seebeck coefficient. For the Heusler phase, $S_1\approx$ $-140\,\mu$V\,K$^{-1}$, and for \ce{Bi_{0.9}Sb_{0.1}}, $S_2\approx -78\,\mu$V\,K$^{-1}$ would be expected. However, instead of two normal distributions centered around those values, the observed distribution appears much more merged with $S$ being significantly under(over)estimated with respect to $S_1$($S_2$). While the underestimation is an artifact from the cold finger effect, inherent to the TPSM and basically all microprobe measurements \cite{ziolkowski2013probing}, the enhanced $S$ of the secondary phase indicates a beneficial interplay between the two components and explains why the integral value of $S$ remains large in the composite (Fig.\,\ref{Fig4}c), despite being short-circuited across the GBs. We note that a similar observation has been made already several years ago by Mikami and Kobayashi in  (\ce{Fe2VAl_{0.9}Si_{0.1}}\,+\,Bi) composites with $zT_\text{max}=0.11$ \cite{mikami2008effect}.

In Fig.\,\ref{Fig4}, we compare the temperature-dependent bulk TE properties of our FVAB$X$ ($X=0,\,20,\,50$) composites over a broad temperature range from 4 to 523\,K. Measurements were performed using different setups in various laboratories at the National Institute for Materials Science (NIMS) in Japan and at TU Wien (TUW) in Austria. Additionally, extensive measurements have also been conducted on a bulk sample of \ce{Bi_{0.9}Sb_{0.1}} synthesized during this study, and the data have been included for comparison. The latter are in excellent agreement with those reported previously (see Fig.\,S13).

Fig.\,\ref{Fig4}a displays the temperature-dependent thermal conductivity $\kappa(T)$. At 200\,--\,300\,K, $\kappa(T)$ increases due to bipolar thermal transport, consistent with the $S(T)$ curves. Most importantly, when \ce{Bi_{1-x}Sb_{x}} is incorporated as a secondary phase, $\kappa(T)$ decreases significantly, which we attribute to the complex microstructural evolution involving microscale \ce{Bi_{1-x}Sb_{x}} GBs with an extremely large acoustic mismatch ($\approx 9\,$THz) with respect to the Heusler matrix, periodic composition fluctuations within the grains, and an enhanced solubility limit of Ta atoms. Moreover, $\kappa(T)$ of the \ce{Bi_{1-x}Sb_{x}} GB network is likely reduced as well compared to the bulk values owing to the ladder-like arrays of stacking fault defects (see Fig.\,\ref{Fig2}f,g) and microscale composition fluctuations (Fig.\,S9), likely inhibiting phonon-driven heat transport along the GBs \cite{isotta2024heat}. 

From Fig.\,\ref{Fig4}b, it is evident that, despite the significant reduction in $\kappa(T)$, electronic transport remains excellent. The temperature-dependent resistivity $\rho(T)$ flattens when \ce{Bi_{0.9}Sb_{0.1}} is incorporated, even resulting in a decrease of $\rho(T)$ at elevated temperatures. The temperature-dependent Seebeck coefficient $S(T)$ only varies moderately in the composites with $S_\text{max}$ shifting to slightly lower temperatures. As mentioned in the previous section, a simple effective-medium theory with parallel conduction along the GB network would result in a sizeable decrease of $S(T)$. The fact that $S$ retains large values, is surprising and unexpected, calling for more advanced theoretical studies of TE transport in composite materials. 
%The $S(T)$ behavior of bulk \ce{Bi_{0.9}Sb_{0.1}} and \ce{Fe2V_{0.95}Ta_{0.1}Al_{0.95}} can be understood by employing a triple- and two-parabolic band model, respectively, which yields a tiny band gap of $E_\text{g} \approx 12$\,meV for the former, with a second conduction band located about 5\,meV below the valence band top, and $E_\text{g} \approx 90$\,meV for the latter. The microscopic band parameters and a sketch of the electronic band structure obtained from the modeling of both materials are shown in the SI.

The above-listed modifications result in an extreme boost of $zT$ by more than a factor of two (see Fig.\,\ref{Fig4}d) up to $zT \approx 0.5$ at 295\,K. This clearly exceeds the predictions of effective-medium theory, which, as demonstrated by Bergman and Levy in their seminal work \cite{bergman1991thermoelectric}, states that $zT$ in composites needs to be always smaller than the largest $zT$ of the individual components, irrespective of the geometry. This implies that the TE properties of the individual components change dramatically in the composite or are subject to reciprocal action of both, allowing for a decoupling of charge and heat transport.
\begin{figure*}[t!]
\newcommand{\setwidth}{0.45}
			\centering
			\hspace*{0cm}
		\includegraphics[width=1\textwidth]{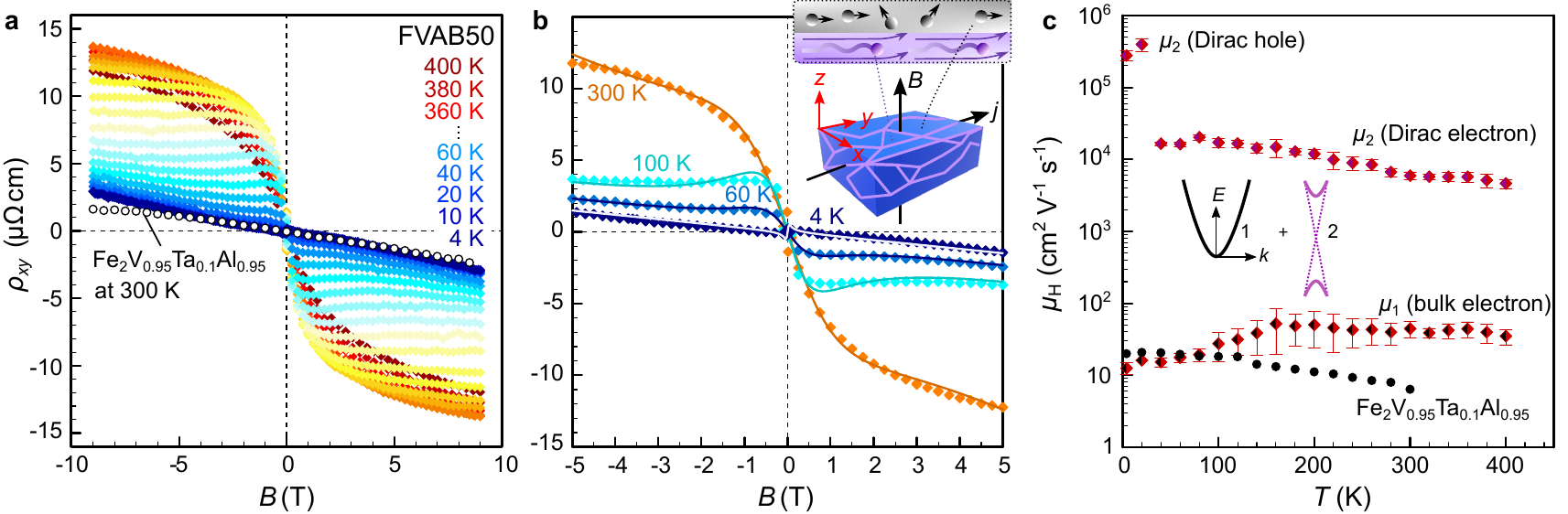}
	\caption{\textbf{$\vert$\,Magneto-transport properties of FVAB50 composite.} \textbf{a},  Field-dependent Hall resistivity of FVAB50 at different temperatures $4\leq T \leq 400\,$K. Room-temperature data of \ce{Fe2V_{0.95}Ta_{0.1}Al_{0.95}} are shown for comparison as black open circles. \textbf{b}, Two-channel transport modelling of field-dependent Hall resistivity. Solid lines are least squares fits. Inset shows a sketch of the Hall effect in FVAB$X$ composites. Highly mobile Dirac-like carriers along the Bi-Sb GB network have a much larger mean free path than the Heusler bulk electrons and are deflected much easier, even in small magnetic fields. \textbf{c}, Temperature-dependent Hall mobility of FVAB50 obtained by modelling the complex field-dependent behavior. Error bars indicate uncertainties of the fit results. The inset shows a sketch of the two transport channels 1) the bulk conduction electrons of the Heusler main phase and 2) the Dirac-like surface states of the Bi-Sb GB network.} 
	\label{Fig5}
\end{figure*}

\vspace*{-0.2cm}
\section*{Analysis of field-dependent magneto-transport}
\vspace*{-0.1cm}
\noindent To further elucidate transport in the best-performing composite sample (FVAB50), we measured the Hall effect in a broad temperature and magnetic field range, 4\,--\,400\,K and $-9$\,T to 9\,T. These results are summarized in Fig.\,\ref{Fig5}. The field-dependent Hall resistivity $\rho_{xy}$, plotted in Fig.\,\ref{Fig5}a for various temperatures displays an extremely large anomalous Hall effect, which even increases with rising temperature up to $\approx 300\,$K, despite the absence of any sizeable magnetization in the sample. On the contrary, the sintered Heusler compound without \ce{Bi_{1-x}Sb_{x}} at the GBs, exhibits a simple linear magnetic field dependence. While the observation of a giant anomalous Hall effect in various topological materials is often ascribed to huge Berry curvatures, emerging from the respective topological band structure features \cite{nayak2016large,manna2018colossal}, we interpret the complex field-dependent curves in Fig.\,\ref{Fig5}a as a two-channel conduction mechanism, where charge carriers can move across the sample either through topologically trivial bulk states of the Heusler grains or through topologically protected surface states of the \ce{Bi_{1-x}Sb_{x}} GB network (see inset Fig.\,\ref{Fig5}b). Fig.\,\ref{Fig5}b shows that the field-dependent behavior of $\rho_{xy}$ from $-5\,$T to 5\,T can be reasonably well described by a simple two-channel transport model (details of the modeling procedure and underlying theory is presented in the SI). The mobilities obtained for the two distinct transport channels are presented in Fig.\,\ref{Fig5}c alongside the values of pristine \ce{Fe2V_{0.95}Ta_{0.1}Al_{0.95}} without topological-insulating GBs. The bulk values of \ce{Fe2V_{0.95}Ta_{0.1}Al_{0.95}} are of the order of 10\,cm$^2$\,V$^{-1}$\,s$^{-1}$. The mobility of the bulk channel, $\mu_1$, extracted from our transport modeling is comparable, especially at low temperatures. The mobility of the Dirac-like surface states, $\mu_2$, associated with the Bi-Sb network, on the other hand, is several orders of magnitude higher up to $3\times 10^5$\,cm$^2$\,V$^{-1}$\,s$^{-1}$ and $2\times 10^4$\,cm$^2$\,V$^{-1}$\,s$^{-1}$ for the Dirac holes and electrons, respectively. As a consistency check, we calculated the temperature-dependent zero-field electrical resistivity from the carrier mobilities $\mu_{1,2}$ and carrier densities $n_{1,2}$ obtained by fitting $\rho_{xy}(B)$ via $\rho_{xx}(0,T)=(e\,n_1 \mu_1 + e\,n_2 \mu_2)^{-1}$, which should match temperature-dependent measurements in Fig.\,\ref{Fig4}b. As shown in Fig.\,S19, there is excellent agreement across the entire temperature range, underscoring the robustness and reliability of the fits.

In summary, the field-dependent Hall effect reveals a significant contribution to electronic transport from the Dirac-like surface states of the \ce{Bi_{1-x}Sb_{x}} GB network, leading to a pronounced anomalous Hall effect, which can be explained by a two-channel transport model. This aligns with the colossal mobilities expected from such topologically robust carriers and the higher surface-area-to-volume ratio in the composite.

\vspace*{-0.2cm}
\section*{Conclusion and Outlook}
\vspace*{-0.1cm}
\noindent Concluding, we demonstrated that incorporating \ce{Bi_{1-x}Sb_{x}} in \ce{Fe2VAl} Heusler compounds can boost the $zT$ compared to both individual materials. This is particularly surprising considering the near-complete immiscibility of both components and their chemical distinctness, which should prevent sizeable interdiffusion and changes to the individual material properties.
%Topological insulators feature a narrow bulk band gap alongside extremely mobile metallic surface states, which are topologically protected from backscattering off non-magnetic defects and impurities. Many topological insulators like \ce{Bi2(Se,Te)3} or \ce{Bi_{1-x}Sb_{x}} are excellent TE materials, which is, however, attributed to their bulk band features (high valley degeneracy of inverted bands due to strong spin-orbit interactions) \cite{toriyama2023band,toriyama2024topological}, and their intrinsically small $\kappa_\text{L}$ arising from the necessity for heavy elements in the crystal structure, and hence, small sound velocities. When TIs are incorporated as low-dimensional GB phases, however, the surface-area-to-volume ratio is significantly enhanced, which can augment the role of their nontrivial surface states. 
Decoupling charge and lattice-driven heat transport in such composites is an auspicious route towards high $zT$, even more so in systems where reducing grain size and alloying strongly compromise carrier mobility, although, a more profound theoretical understanding of charge and heat transport in composites is required to optimally design and choose the best candidates.

In this study, we achieved heavily reduced $\kappa_\text{L}$, and simultaneously high $\mu_\text{W}$. To provide a broad comparison with other material classes for the latter, we downloaded all available TE property data from the \textsf{\textit{Starrydata2}} open web database \cite{katsura2019data}, which, as of July 2024, contains TE data from 8,961 different papers and 52,020 different samples. We then calculated $\mu_\text{W}$ for those samples, where both $S(T)$ and $\rho(T)$ are reported. Fig.\,S22 shows that, near room temperature, $\mu_\text{W}$ of the best composite sample from this work surpasses all other reported $n$-type semiconductors.

To further elevate the performance of \ce{Fe2VAl} systems, broad screening of secondary phases needs to be done; especially other topological insulators like \ce{Bi2Se3} could be considered. Additionally, one has to think about strategies to increase and reliably tune the volume fraction. Lastly, it is crucial to identify promising $p$-type compounds with competitive $z$. Since $p$-type \ce{Fe2VAl} compounds inherently show much smaller Seebeck coefficients, this can only be realized via band engineering of the valence band electronic structure \cite{nishino2019effects,parzer2022high}. Only then can competitive \ce{Fe2VAl}-based modules be realized, which could substitute the long-reigning \ce{Bi2Te3} systems. The present study suggests that, by proper GB engineering, \ce{Fe2VAl} Heusler alloys may indeed bear the potential to rival state-of-the-art \ce{Bi2Te3} and \ce{Mg3Bi2} semiconductors. High-performance modules entirely based upon \ce{Fe2VAl} alloys could open a new era for near-ambient applications, as these systems excel in terms of cost-effectiveness, excellent recyclability and simpler device structures. Moreover, they exhibit superior mechanical, thermal and chemical long-term stability, factors that are becoming increasingly recognized as essential assets for realizing widespread thermoelectric technology.

%\end{linenumbers}
%\bibliographystyle{apsrev4-2}
%\bibliography{bibliography_composite}
%

\vspace*{-0.2cm}
\footnotesize{\section*{Acknowledgments}
\vspace*{-0.1cm}
\noindent Research in this paper was financially supported by the Japan Science and Technology Agency (JST) program MIRAI, JPMJMI19A1 as well as by the Lions Club Wien St. Stephan.}

\end{document}